\documentstyle[epsfig]{aipproc2}

\newcommand{\Msolhae}{\mbox{M$_\odot$}}
\newcommand{\yrhae}{\,{\rm yr}}
\newcommand{\Mpchae}{\,{\rm Mpc}}
\newcommand{\kmshae}{\mbox{km\,s$^{-1}$}}

\begin{document}

\title{Supermassive black holes as sources for LISA}

\author{Martin G. Haehnelt}
\address{Institute of Astronomy,  Madingley Road, 
Cambridge CB3 0HA, UK.}

\maketitle

\begin{abstract}
I briefly discuss some  issues relevant for the formation of 
supermassive black holes and give estimates of the event rates 
for the emission of gravitational waves by coalescing supermassive
black hole binaries.  I thereby use models which take into 
account recent improvements in our knowledge of galaxy and 
star formation in the high-redshift universe. Estimated event 
rates range from a few to a hundred per year. Typical 
events will occur at redshift three or larger in galaxies 
lying at the (very) faint end of the luminosity function at these 
redshifts. 
\end{abstract}

\section*{Introduction}

Supermassive black holes (SMBH's) are amongst the prime targets for LISA 
and  LISA  will be primarily sensitive to events 
involving SMBH's  in the mass range $10^{4-6} \Msolhae$ over 
a wide range range in redshift \cite{hae:lis98,hae:hae94,hae:ree97}
(see also the contributions by Blandford 
and Sigurdsson these proceedings). 
The evidence for the existence of SMBH's  more massive than that 
has been steadily  increasing over the last 
years. The two most convincing cases are currently our own galactic 
centre and NGC4258 \cite{hae:gen97,hae:wat94,hae:miy95}. 
In both cases the inferred deep potential wells and 
high mass densities leave little room for alternative explanations
other than the presence of a SMBH \cite{hae:mao95}.  
Our best  estimate of the overall mass density 
in black holes still comes from the integrated flux emitted   
by optically bright QSO's which are generally believed 
to be predominantly powered by SMBH's.  
From this  a present-day black hole mass density of 
$\sim 1.5\times 10^5 \Msolhae \Mpchae^{-3}$ was inferred  
\cite{hae:sol82,hae:cho92} which  corresponds to about 
$5\times 10^{7} \Msolhae$ per $L_{*}$  galaxy. This estimate  
has recently been complemented by an investigation of 
a large  sample of black hole masses for 
nearby galaxies which gives a factor three to ten higher 
value suggesting that the mass of the typical black hole 
of a galaxy could be as high as 0.6\%  of the  stellar mass 
contained in its bulge \cite{hae:mag98,hae:mar98}
(see also Richstone these proceedings for a discussion).

\section*{The formation of  supermassive  black holes}

A variety of more or less detailed scenarios has been 
suggested for the formation of SMBH's
(see Rees 1984 \cite{hae:ree84} for a review). These scenarios 
involve  one or several of the following basic processes 
leading to the concentration of mass,  
\begin{itemize}
\item
{the dynamical evolution of a dense cluster of stellar objects,}  
\item{the build-up of a supermassive black hole by merging 
of supermassive black holes of smaller mass,}  
\item{and the viscous evolution and eventual collapse of 
a self-gravitating gaseous object (barred or non-barred disc, 
supermassive star).} 
\end{itemize}

While all these processes will occur the first two have serious 
difficulties  when it comes to explain the existence of typical 
SMBH's observed in  optically bright QSO's or nearby galaxies. 
I will briefly discuss the options one by one. The main problem 
with the ``stellar route'' to a SMBH  is the rather  long  dynamical 
relaxation timescale $t_{\rm rel} \sim  3\times 10^{10} 
\, v_{300}^{-3}\, N_{8}^2\, m_{*}\, \log{(N/2)}^{-1}\yrhae $,  
where $v_{300}$ is the 1D-velocity dispersion of the stellar cluster 
and $N_{8}$ is the number and $m_*$
the mass of  stellar objects.  Typical 
masses of black holes  powering high-redshift QSO's are 
$\sim 10^{8-9} \Msolhae$. As the stellar cluster from which these could
form has to be even more massive the relaxation timescale is 
prohibitively long.  It is generally very difficult to concentrate 
the mass in an  efficient manner once the gas has fragmented into 
stars.  

\begin{figure}[t!] 
\centerline{\epsfig{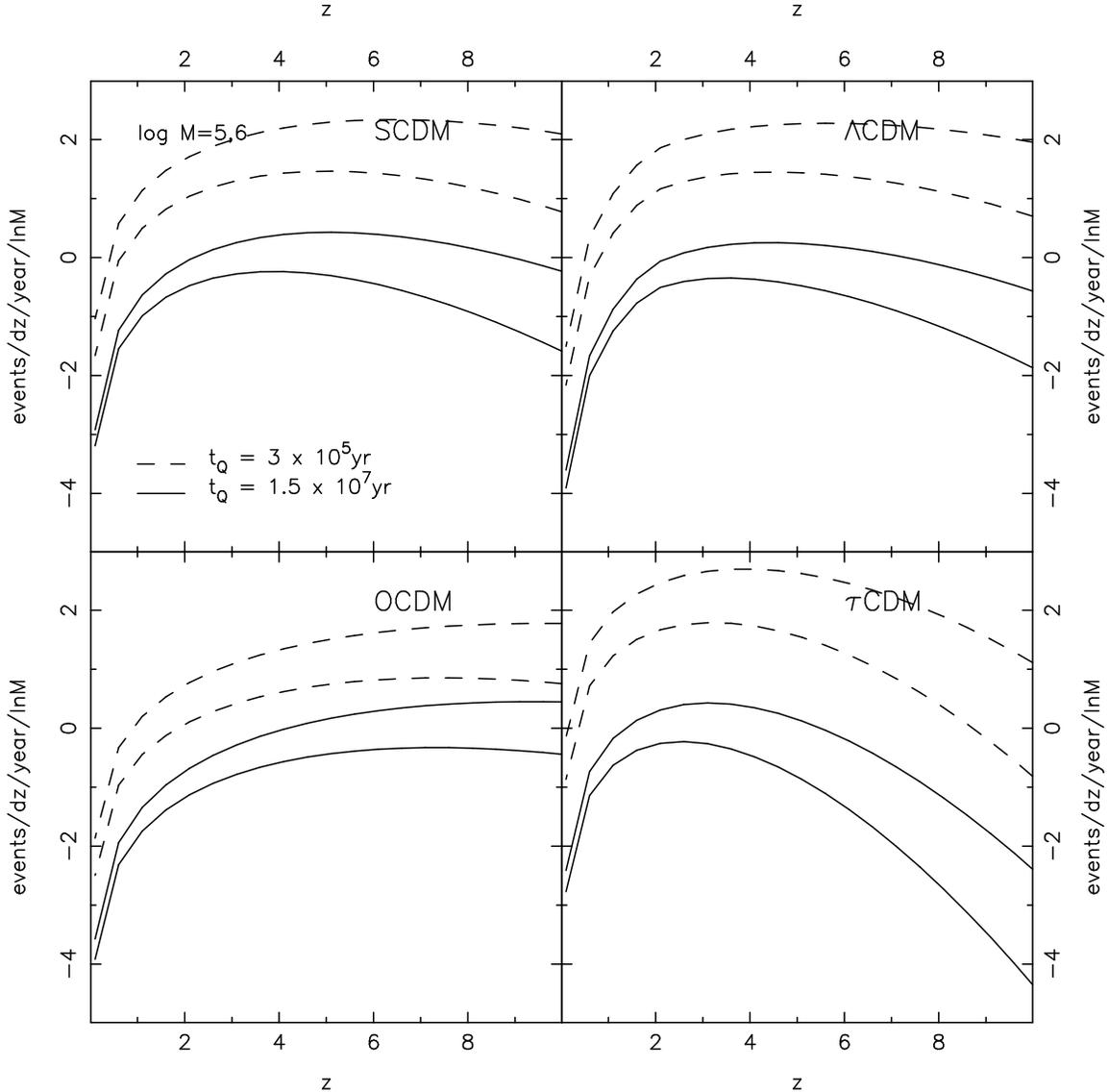}}
\vspace{10pt}
\caption{Event rates for the emission of gravitational waves involving 
supermassive black holes above a certain mass as indicated on the
plot. The four panels are for different variants of the
cold-dark-matter cosmogony (see table 1 for parameters). 
One event  per newly-formed dark matter halo is  assumed.
Dashed and solid curves are for two different QSO lifetimes  
used to calibrate the models as indicated in the upper left panel. 
The upper and lower curves are for $10^5\Msolhae$ and  
$10^6\Msolhae$, respectively.}

\label{foobar:fig1}
\end{figure}

The ``merging scenario'' for the formation of SMBH's has  
become increasingly attractive with the accumulating
observational evidence for  the  hierarchical build-up of 
large galaxies predicted 
by CDM-like structure formation scenarios. In hierarchical 
cosmogonies typical  present-day galaxies have  formed by merging of 
about ten smaller galaxies between redshift three and now and each of 
these ``progenitors'' will have formed from even smaller sub-units 
at  higher redshift. When these galaxies merge the putative black
holes at their centre will generally merge as well 
\cite{hae:beg80}. If merging were 
indeed the dominant process for the build-up 
of the mass of SMBH's the problem of the formation of present-day 
SMBH's could  be deferred to the problem of the 
formation of much  smaller mass SMBH's in galactic sub-units 
at very high redshift \cite{hae:hai98}. These would, 
however, have to form with the high efficiency inferred from the large 
black-hole mass to stellar-bulge mass ratio of  present-day galaxies
in small protogalactic clumps with shallow potential wells.   
As argued by Haehnelt\& Rees \cite{hae:hae93} and  Haehnelt, Natarajan 
\& Rees (HNR98) \cite{hae:hae98} the black-hole formation efficiency 
should be larger in the deeper  potential wells of the larger 
galaxies forming around redshift three making this an unlikely 
but not impossible option.  

This leaves the last option where most of the mass is accreted 
in  gaseous form  --- the fastest and most efficient way to 
concentrate mass in SMBH's.  
The typical timescale for the concentration 
of the mass will be $10^{7-8} \yrhae$, 
the dynamical  time scale  of the galactic nucleus. The 
viscosity is mainly due to  gravitational instabilities 
inevitably present in a self-gravitating  disc. 
There is  no fundamental limit for the efficiency of this process   
but the concentration of the gas has to compete with the consumption 
of the gas by star formation which will occur on a similar timescale. 
Unfortunately the expected gradual build-up of the mass is not likely 
to produce gravitational waves efficiently  \cite{hae:ree97}. It will, however, 
be accompanied by the frequent merging characteristic 
for  hierarchical cosmogonies and  probably also by the infall of 
SMBH's of  intermediate mass which may form via the stellar 
route in the central star cluster at the galactic nucleus.
In the next  section we will 
give estimates of the expected  event rates and briefly 
discuss some of the uncertainties involved.

\section*{Event rates in hierarchical structure \\formation scenarios} 

\begin{figure}[t!] 
\centerline{\epsfig{file=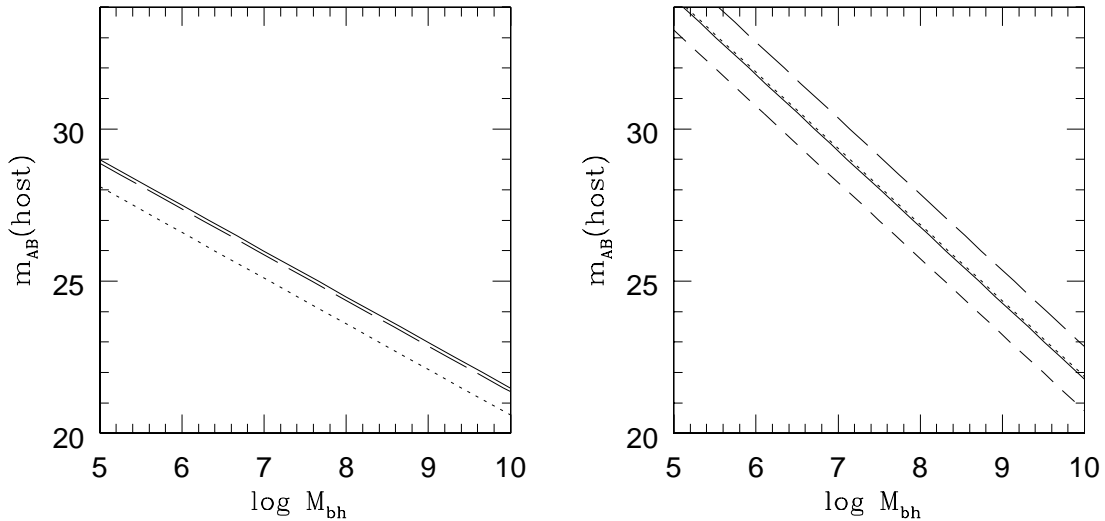,width=6.5in}}
\vspace{-8.0cm}
\caption{Typical apparent brightness of host galaxies at z=3. The
right and left panel assume a QSO lifetime of $1.5\times 10^{7} \yrhae$
and  $3\times 10^{5} \yrhae$, respectively. Different lines are for 
different cosmological models.}
\label{foobar:fig2}
\end{figure}

Structure in the universe is  generally believed to originate from 
small density fluctuations of some sort of collisionless dark matter
(DM). These fluctuations are assumed to be Gaussian distributed  and 
can be  specified  by their spatial power spectrum. The 
dynamical evolution of the dark matter, the  mass function of 
collapsed DM  halos and their merging rates can be 
reliable predicted once the  cosmological model is specified. 
The main difficulty in a comparison with astronomical data is to
predict the distribution of gas and stars relative to that of the 
dark matter.  Here we would like to know 
the space density and  merging rate of SMBH's
to  estimate  the  gravitational wave event rates 
due to coalescing SMBH binaries.
Haehnelt \& Rees and HNR98 have demonstrated that with simple relations 
between DM halo and black hole mass  observed high-redshift galaxies
and QSO's can be explained reasonably well. The models proposed 
in HNR98 predict the ``formation rate'' of newly-formed DM halos and 
thus allow us to get a crude estimate of the formation rate of 
SMBH's as function of their mass. 
However, these estimates depend on the uncertain lifetime of
optically bright QSO's as they are calibrated by a comparison 
with the QSO space density.  
The typical formation of a DM halo involves the merging
of several DM halos and this formation should be 
accompanied by one or several merging events of SMBH's. Figure 1 
gives estimate of event rates and makes the rather 
conservative assumption that each newly-formed DM 
halo produces one gravitational wave event. The 
solid curves are for a lifetime of 
$1.5 \times 10^7\yrhae $ and assume a non-linear relation
between black hole and DM halo mass as discussed by HNR98. Typical
event rates are one per year about the same as those 
obtained by Haehnelt (1994)  \cite{hae:hae94} for similar assumptions.  
The four panels are for different  cosmogonies which span 
the range of currently viable models. The relevant parameters 
are given in table 1. 
Obviously the numbers  have to  be convolved with the sensitivity 
curve of LISA.  LISA should  detect the coalescence of equal mass 
binaries of $10^{5-6}\Msolhae$  out to maybe redshift ten. Even though 
at high redshift a sufficient signal-to-noise ratio will probably 
only be achieved for considerably less than a year. The coalescence 
of unequal mass SMBH binaries will only be detectable at significantly
lower redshift.   One should also keep in mind
that little is known observationally about black holes of such small
mass especially at high redshift and that 
the predictions rely on an extrapolation of the models 
from the typical black hole of mass $10^{8} \Msolhae$ or
larger observed around redshift three. 
Furthermore at low redshift (below $z \sim 2$)  the numbers in
Figure 1 are likely to  underestimate the merging rate of SMBH's as no
attempt was made to model the late merging of galaxies in DM halos 
which formed at earlier times. 

As demonstrated  by the dashed curves event rates would be about a factor
thirty  higher if the lifetime of QSO's were as short  as  
$3 \times 10^5 \yrhae$. In this  case  a linear relation between 
DM halo mass and black hole mass is required \cite{hae:hae98}. The lifetime of 
the QSO's  will also affect the  predicted host-galaxy luminosity
and the clustering properties of the QSO's.  Constraints on both of
these should be soon improved by the planned new large 
QSO surveys (2DF, SLOAN)  and the uncertainty in the QSO lifetime 
reduced or even removed. 
  
Figure 2 shows the predicted  apparent brightness of typical  
host-galaxies at $z=3$ as a function of black hole mass 
for the cosmological models
in  Fig.~1 . The left panel is for 
the long lifetime and the right panel for the short lifetime. Note 
that especially for short QSO lifetimes  most of the 
predicted coalescences should occur in extremely faint galaxies.

\begin{table}
\caption{The model parameters of the CDM variants explored:
$\sigma_8$ is the  {\it rms} linear overdensity in spheres of radius    
$8\,h^{-1}\,{\rm Mpc}$ and $\Gamma$ is a shape parameter 
for CDM-like spectra. h is the Hubble constant
in units of the $100\kmshae$ and $\Omega_0$ and  ${\Omega_{\Lambda}}$   
are the total energy  density and that due to a cosmological constant,
respectively. }
\begin{tabular}{|l|l|l|l|l|l|}
\hline
${\rm
MODEL}$&${\sigma_8}$&${h}$&${\Omega_0}$&${\Omega_{\Lambda}}$&${\Gamma}$\\
\hline
${\rm SCDM}$ & ${0.67}$ & ${0.5}$ & ${1.0}$ & ${0.0}$ & ${0.5}$ \\
\hline
${\rm OCDM}$ & ${0.85}$ & ${0.7}$ & ${0.3}$ & ${0.0}$ & ${0.21}$ \\
\hline
${\rm \Lambda CDM}$ & ${0.91}$ & ${0.7}$ & ${0.3}$ & ${0.7}$ & ${0.21}$ \\ 
\hline
${\rm \tau CDM}$ & ${0.67}$ & ${0.5}$ & ${1.0}$ & ${0.0}$ & ${0.21}$ \\
\hline
\end{tabular}
\end{table}

\section*{Discussion}

The existence of supermassive black holes in a major fraction of all 
galaxies seems firmly established. Most of the mass in these
SMBH's  will have found its way beyond the event horizon 
in the gas-rich nuclei of the progenitors of present-day  
galaxies at high redshift. The emission of gravitational waves 
from coalescing supermassive binary black holes formed during the merger 
of such proto-galaxies should occur frequently enough 
to be detected during the lifetime of LISA. 
Typical  events should occur in proto-galaxies at the 
(very) faint end of the luminosity function. 
The event rates are expected to increase 
with redshift with a rather broad peak at
redshift three or larger and a slow decline at higher redshift
The details of this decline are very uncertain and  depend 
strongly on how efficiently small black holes form in shallow 
potential wells. The biggest uncertainty, however, is 
the number of detectable coalescences  per newly-formed halo. 
Each new halo will be formed by the  merging of a number of 
smaller halos each of which will contain one or more  
SMBH's. 
How many coalescences  occur will depend on whether  
the black hole binary formed in one 
merger  has already coalesced  when the next black hole sinks to 
the centre. Otherwise sling-shot ejection  is possible
 \cite{hae:beg80}. Furthermore the black holes
will be embedded in a nuclear star cluster formed from gas on its 
way into the SMBH.  These star cluster will not 
contribute much to the total mass in SMBH's. 
Nevertheless,  they still  could form
with high velocity dispersion and  black 
holes of 100 to $10^4 \Msolhae$ might build up 
efficiently by the  coalescence of stars 
 \cite{hae:qui90}.
At lower redshifts  these  would also be detectable by LISA 
when the coalesce with the central SMBH. 

I finally conclude, that even a pessimist who assumes  a rather
long QSO lifetime and only one binary coalescence per newly-formed  
halo should  expect a couple of SMBH binary
coalescences during the  lifetime of LISA while an optimist 
might expect to see up to several hundred of these exciting 
events. 
   
\section*{Acknowledgments}
I would like to thank Martin Rees for many stimulating  discussions  
on the issues discussed here.

\end{document}